\documentclass[aps,prl,twocolumn,superscriptaddress]{revtex4}
\usepackage{graphicx}
\usepackage{amssymb}
\usepackage{amsmath}
\usepackage{graphicx}
\usepackage{epsfig}
\usepackage{color}
\usepackage{bm}

\begin{document}

\title{A unified theory of ferromagnetic quantum phase transitions in heavy fermion metals}
\author{Jiangfan Wang}
\affiliation{Beijing National Laboratory for Condensed Matter Physics, Institute of Physics,
Chinese Academy of Science, Beijing 100190, China}
\author{Yi-feng Yang}
\email[]{yifeng@iphy.ac.cn}
\affiliation{Beijing National Laboratory for Condensed Matter Physics,  Institute of Physics, 
Chinese Academy of Science, Beijing 100190, China}
\affiliation{School of Physical Sciences, University of Chinese Academy of Sciences, Beijing 100190, China}
\affiliation{Songshan Lake Materials Laboratory, Dongguan, Guangdong 523808, China}
\date{\today}

\begin{abstract}
Motivated by the recent discovery of a continuous ferromagnetic quantum phase transition in CeRh$_6$Ge$_4$ and its distinction from other U-based heavy fermion metals such as UGe$_2$, we develop a unified explanation of their different ground state properties based on an anisotropic ferromagnetic Kondo-Heisenberg model. We employ an improved large-$N$ Schwinger boson approach and predict a full phase diagram containing both a continuous ferromagnetic quantum phase transition for large magnetic anisotropy and first-order transitions for relatively small anisotropy. Our calculations reveal three different ferromagnetic phases including a half-metallic spin selective Kondo insulator with a constant magnetization. The Fermi surface topologies are found to change abruptly between different phases, consistent with that observed in UGe$_2$. At finite temperatures, we predict the development of Kondo hybridization well above the ferromagnetic long-range order and its relocalization near the phase transition, in good agreement with band measurements in CeRh$_6$Ge$_4$. Our results highlight the importance of magnetic anisotropy and provide a unified theory for understanding the ferromagnetic quantum phase transitions in heavy fermion metals. 

\end{abstract}

\maketitle

\section{Introduction}
Heavy fermion or Kondo lattice materials are prototype strongly correlated electron systems that exhibit rich ground states due to competing interactions \cite{YangPNAS2017,review_si2010}. An indirect Ruderman-Kittel-Kasuya-Yosida (RKKY) interaction is typically induced and oscillates with distance between neighboring Kondo impurities. Its resulting ferromagnetic (FM) or antiferromagnetic (AFM) long-range orders can often be suppressed by pressure, magnetic field, or chemical substitution and tuned to a heavy Fermi liquid (HFL) with an enhanced effective mass of Landau quasiparticles due to the Kondo screening by conduction electrons. However, unlike AFM Kondo lattices, whose magnetic quantum phase transitions (QPTs) and associated quantum criticality have been intensively studied in the past decades \cite{review_Wirth,review_Coleman}, the FM Kondo lattices are relatively less explored \cite{Yashar1D, Si_PNAS, ZhangGuangMing2010,Peters2012,Golez2013}. It has been believed that FM QPT in clean metallic systems should typically be of first order if not interrupted by other instabilities \cite{BrandoRMP2016} because of non-analyticities in the free energy induced by certain soft modes related to the  Fermi surfaces \cite{Kirkpatrick1999,Kirkpatrick2012}. Evidences may be found from experiments on many U-based intermetallics such as UGe$_2$ \cite{UGe2_review}.

The situation, however, has been changed lately when a pressure-tuned FM quantum critical point (QCP) and the associated strange metallic behaviors were observed in the stoichiometric Kondo lattice compound, CeRh$_6$Ge$_4$ \cite{CeRhGe_nature}. Similar observations had been reported previously in the non-stoichiometric compound YbNi$_4$(P$_{1-x}$As$_x$)$_2$ \cite{YbNiP_science}, but might be naively attributed to disorder effects. These observations violate the prevailing wisdom and have stimulated rapid theoretical progress and debates on the underlying mechanism \cite{CeRhGe_nature, Kirkpatrick2020}. Analogous to the AFM quantum criticality, a Kondo breakdown scenario has also been predicted across the FM QCP \cite{CeRhGe_nature}. But this picture cannot explain the latest observations of angle-resolved photoemission spectroscopy (ARPES) \cite{ARPES_PRL2021} and ultrafast optical spectroscopy \cite{pump_PRB2021}, where strong hybridization effects have been detected at finite temperatures above the FM transition.

Compared to CeRh$_6$Ge$_4$, the intensively studied U-based compound, UGe$_2$, exhibits very different ground state properties under pressure \cite{UGe2_review}. At low pressures, it has two different FM phases with strong (FM2) and weak (FM1) spin polarizations, respectively, and enters a HFL upon further increasing pressure. Both FM2-FM1 and FM1-HFL transitions are of first order \cite{Pfleiderer2002}, with drastic changes of Fermi surfaces reported by quantum oscillation experiments \cite{Terashima2001, Haga2002}. A superconducting dome occurs inside the FM phase and the maximal superconducting transition temperature occurs right at the FM2-FM1 transition point \cite{UGe2_review}.

Given these distinct properties, one may doubt if there should exist a single mechanism for both compounds without resorting to microscopic details. In this work, we propose such a unified theory and show that their very different properties can in fact be explained solely based on lattice anisotropy. The Ce atoms of CeRh$_6$Ge$_4$ exhibits a quasi-one-dimensional structure \cite{CeRhGe_nature}, while the nearest neighbor U-U separations in UGe$_2$ are more isotropic \cite{Oikawa1996}. Such structural anisotropy on atomic distances may have a sensitive influence and cause a magnetic anisotropy in the RKKY interaction, as reflected in an anisotropic Kondo-Heisenberg model. We will solve the model using an improved Schwinger boson large-$N$ approach taking into consideration the spatial correlation effect. The method has lately been applied to the AFM Kondo lattice \cite{Wang2021}, yielding a global phase diagram and a unified explanation of the AFM QCP in YbRh$_2$Si$_2$ and the quantum critical non-Fermi liquid phase (a metallic spin liquid) recently observed in the frustrated Kondo lattice CePdAl \cite{Zhao2019}. Here we further develop this approach for the FM phase, predict a full phase diagram of the anisotropic FM Kondo lattice, and give a unified explanation of the rather complicated and qualitatively different behaviors of CeRh$_6$Ge$_4$ and UGe$_2$. Our methods and results are reported in detail in the following sections. 

\section{Methods}

We start with the Hamiltonian:
\begin{eqnarray}
H=\sum_{\left\langle ij\right\rangle}t_{ij}c_{ia\alpha}^\dagger c_{ja\alpha}+J_K\sum_{i}{\bf S}_{i}\cdot {\bf s}_{i}-\sum_{i,\delta}J_{H}^{\delta}{\bf S}_{i}\cdot {\bf S}_{i+\delta},
\label{H}
\end{eqnarray}
where $c_{ia\alpha}^\dagger $ creates a conduction electron of spin $\alpha$ and channel (orbital) $a = 1,2,\cdots,K$ on site $i$ of a square lattice (summations over repeated indices are implied), ${\bf s}_{i}=\frac{1}{2}\sum_{a\alpha\beta}c_{ia\alpha}^\dagger \bm{\sigma}_{\alpha\beta}c_{ia\beta}$ is its spin operator, and ${\bf S}_{i}$ denotes the local spin formed by either Ce 4$f$-electrons or U 5$f$-electrons beyond the Hill limit \cite{Hill1970}. $\delta=x, y$ denotes the unit vectors along the two spatial directions. In principle, the Kondo couping will induce an anisotropic RKKY interaction if the electron hopping $t_{ij}$ is anisotropic. Here, we apply directly an anisotropy on the FM Heisenberg interaction ($J_H^x\neq J_H^y >0$) while keeping the electron hopping isotropic to emphasize the effect of magnetic anisotropy alone.

To deal with FM and Kondo effect on the same basis, we use the Schwinger boson representation of local spins, $\bm{S}_i=\frac{1}{2}\sum_{\alpha\beta}b_{i\alpha}^\dagger \bm{\sigma}_{\alpha\beta}b_{i\beta}$, together with the constraint $n_b(i)\equiv \sum_{\alpha} b_{i\alpha}^\dagger b_{i\alpha}=2S$, and perform large-$N$ calculations by extending the spin group SU(2) to SU($N$) with $\alpha=1,\cdots, N$ and a fixed ratio $\kappa\equiv 2S/N=K/N$ from the perfect Kondo screening condition \cite{Coleman-SWB-2006}. The constraint is implemented by a Lagrange multiplier, $\lambda \sum_i(n_b(i)-2S)$. The Kondo and Heisenberg terms can be decomposed by two auxiliary fields:
\begin{eqnarray}
J_K {\bf S}_{i}\cdot {\bf s}_{i} &\rightarrow& \frac{1}{\sqrt{N}}b_{i\alpha}^\dagger c_{ia\alpha}\chi_{ia}+H.c.+\frac{|\chi_{ia}|^2}{J_K}, \notag \\
-J_H^{\delta} \bm{S}_i\cdot \bm{S}_{i+\delta} &\rightarrow & \Delta_{\delta} b_{i+\delta,\alpha}^\dagger b_{i\alpha}+H.c.+\frac{N|\Delta_{\delta}|^2}{J_H^\delta},
\label{eq:HS}
\end{eqnarray}
where $\chi_{ia}^\dagger$ creates a spinless fermion with positive charge called holon, and $\Delta_{\delta}=-J_H^{\delta}\sum_\alpha \left\langle b_{i\alpha}^\dagger b_{i+\delta, \alpha}\right\rangle /N$ denotes the hopping amplitude of bosonic spinons along the $\delta$-direction. We will use the mean-field ratio $\Delta_x/\Delta_y$ to measure the degree of magnetic anisotropy. 

In the large-$N$ limit, one only needs to consider the one-loop self-energies of spinons and holons:
\begin{eqnarray}
\Sigma _{b}({\bm{p}}, i\nu_n)&=&-\frac{\kappa}{\beta\mathcal{V}} \sum_{\bm{k}m}g_c(\bm{p}-\bm{k}, i\nu_n-i\omega_m ) G_{\chi }(\bm{k}, i\omega_m ), \notag \\
\Sigma _{\chi }(\bm{p}, i\omega_m)&=&\frac{1}{\beta\mathcal{V}}\sum_{\bm{k}n}g_c(\bm{k}-\bm{p}, i\nu_n -i\omega_m ) G_{b}(\bm{k}, i\nu_n ),
\label{eq:SelfE}
\end{eqnarray}
where $g_c(\bm{p},i\omega_m)=(i\omega_m-\varepsilon_{\bm{p}}^c)^{-1}$ is the bare Green's function of conduction electrons with dispersion $\varepsilon_{\bm{p}}^c$, $G_{b}$ and $G_{\chi }$ are the self-consistent Green's functions of spinons and holons, $\omega_m$ ($\nu_n$) are the fermionic (bosonic) Matsubara frequencies, $\beta$ is the inverse temperature, and $\mathcal{V}$ is the total number of lattice sites. 

The Green's functions satisfy:
\begin{eqnarray}
G_b(\bm{p},i\nu_n)&=&\frac{1}{i\nu_n-\varepsilon_{\bm{p}}^b-\Sigma_b(\bm{p},i\nu_n)}, \notag \\
G_\chi(\bm{p}, i\omega_n)&=&\frac{1}{-J_K^{-1}-\Sigma_\chi(\bm{p},i\omega_n)-\frac{b^2}{i\omega_n+\varepsilon_{-\bm{p}}^c}}, \notag \\
G_{c}^\alpha(\bm{p}, i\omega_n)&=&\frac{1}{i\omega_n-\varepsilon_{\bm{p}}^c+\frac{b^2 \delta_{\alpha,1}}{-J_K^{-1}-\Sigma_\chi(-\bm{p},-i\omega_n)}},
\label{eq:GF}
\end{eqnarray}
where $\varepsilon_{\bm{p}}^b\equiv \lambda-2(\Delta_x\cos p_x+\Delta_y \cos p_y)$ is the bare spinon dispersion. The FM long range order is accounted for by the condensation of bosonic spinons on one of its spin component (chosen to be $\alpha=1$ here), $\left\langle b_{i,1}\right\rangle=\sqrt{N}b\neq 0$, so that $\left\langle S_z\right\rangle\equiv N^{-1}\sum_\alpha\text{sgn}(\frac{1+N}{2}-\alpha)\langle b_{i\alpha}^\dagger b_{i\alpha}\rangle=b^2$ \cite{Lebanon2006}. 

In this formalism, the Kondo effect at large $J_K$ is signaled by the negative effective Kondo coupling (or holon energy level), $J_K^*(\bm{k})^{-1}\equiv  J_K^{-1}+\text{Re}\Sigma_\chi(\bm{k},0)<0$, which energetically favors formation of Kondo bound states. It then leaves an energy gap around the Fermi energy for both holons and spinons, producing a large electron Fermi surface due to the generalized Luttinger sum rule \cite{coleman2005sum, Wang2021}.  As one increases $J_H^{\delta}$, the spinon (holon) gap shrinks and finally vanishes at some point where the bosonic spinons start to condense due to the development of long-range FM order. Inside the FM phase, the condensation (magnetization) provides an effective hybridization field between the conduction electrons ($c_{ia,\alpha=1}$) and the holons, as may be directly seen from the decomposed Kondo term in Eq. \eqref{eq:HS}.

In practice, Eqs. (\ref{eq:SelfE})(\ref{eq:GF}) are solved self-consistently under two constraints corresponding to minimizing the free energy with respect to $\lambda$ and $b$. Previous studies of the Kondo lattice models using Schwinger boson approach have only focused on the paramagnetic phase and adopted a local (or momentum-independent) approximation for the self-energies \cite{Yashar1D,Komijani2019,wang2019quantum}. Here we use the fast Fourier transform of the self-consistent equations and solve the nonlocal (or momentum-dependent) self-energies efficiently in the coordinate space \cite{Wang2021}. More details can be found in Supplementary Materials. 

\begin{figure}
\centering\includegraphics[scale=0.435]{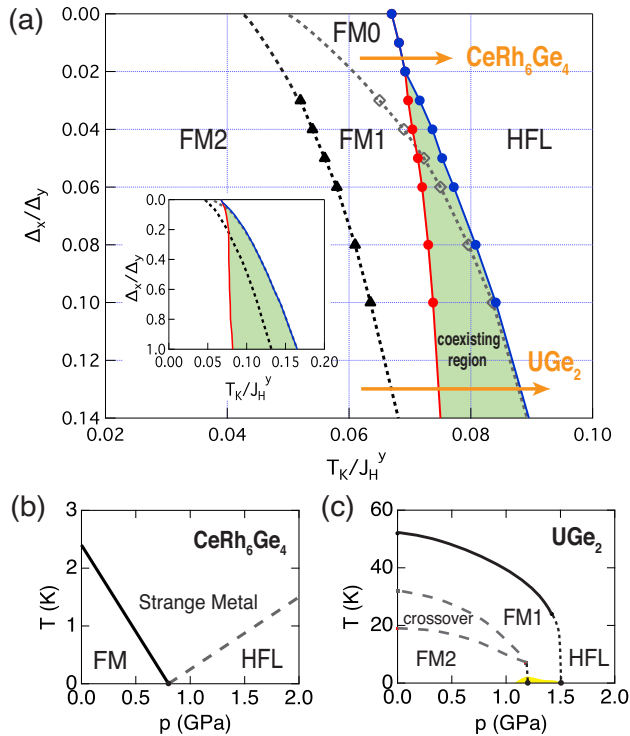}
\caption{\textbf{Phase diagram of the anisotropic ferromagnetic Kondo lattice.} (a) Zero temperature phase diagram of the model described by Eq. (\ref{H}). The heavy Fermi liquid (HFL) phase at large $T_K/J_H^y$ ends at the red solid line, while the ferromagnetic (FM) phases at small $T_K/J_H^y$ end at the blue solid line. Between the red and blue lines (light green area), HFL and FM solutions coexist.  The three different FM phases, FM0, FM1 and FM2, are separated by two continuous transition lines (the black and grey dashed lines). The two orange arrows correspond to increasing pressure of CeRh$_6$Ge$_4$ and UGe$_2$. The inset shows the phase diagram within the  full range of anisotropy. (b)(c) The experimental temperature-pressure phase diagram of CeRh$_6$Ge$_4$ and UGe$_2$, respectively. Black lines are continuous transitions, short dashed lines are first-order transitions, and long dashed lines are crossover lines. The small yellow region of (c) denotes the superconducting phase. The experimental phase diagrams are reproduced based on Refs. \cite{CeRhGe_nature} and \cite{UGe2_review}.}
\label{fig:Global}
\end{figure}

\section{Results}

We first summarize the predicted zero temperature phase diagram in Fig. \ref{fig:Global}a in terms of the anisotropy ratio $\Delta_x/\Delta_y$ and the Doniach ratio $T_K/J_H^y$, where $T_K=De^{-2D/J_K}$ is the characteristic Kondo energy scale and $D$ is the electron half bandwidth. The HFL phase exists on the right hand side of the red line (larger $T_K/J_H^y$), and the FM phases appear on the left hand side of the blue line (smaller $T_K/J_H^y$). The two lines merge together for large anisotropy (small $\Delta_x/\Delta_y$), indicating a continuous quantum phase transition and a single FM QCP. While for smaller anisotropy (larger $\Delta_x/\Delta_y$), there is a coexisting region (light green area) between the two lines, indicating a first-order transition between FM and HFL. It should be mentioned that the Schwinger boson large-$N$ theory suffers from the pathology of predicting artificial first-order transitions \cite{fczhang2002pathology}, which can be alleviated via introducing a biquadratic interaction term, $-\zeta\sum_{i,\delta}J_H^\delta (\bm{S}_i\cdot \bm{S}_{i+\delta})^2$ \cite{Yashar1D,CeRhGe_nature,Wang2021}. In our case, such a term with a reasonable choice of $\zeta$ only slightly shifts the tricritical point from $\Delta_x/\Delta_y\approx 0.02$ to $0.03$ (see Supplementary Materials) and has no effect on the overall picture and qualitative properties.   

The FM phase is further divided by another two continuous transition lines  into three different regions: FM0, FM1, and FM2. The FM1-FM0 transition line merges with the FM phase boundary (the blue line) at large $\Delta_x/\Delta_y$, so the FM0 phase only exists at small $\Delta_x/\Delta_y$ (large anisotropy) and has the smallest magnetization among the three FM phases. It can be suppressed continuously by increasing $T_K/J_H^y$ and becomes a HFL. As shown in Fig. \ref{fig:Global}b, FM0 may correspond to the observed FM phase of CeRh$_6$Ge$_4$ at ambient pressure, which also exhibits a severely reduced magnetic moment (0.28$\mu_B$/Ce \cite{CeRhGe_nature} as compared to the estimated 1.28$\mu_B$/Ce from crystalline electric field analysis \cite{CeRhGe_CEF2021}) and is turned continuously into a HFL through a FM QCP by applying pressure \cite{CeRhGe_nature}. The leftmost FM2 phase has the largest magnetization, which decreases gradually with increasing $T_K/J_H^y$, while the intermediate FM1 phase is featured with a constant magnetization insensitive to the value of $T_K/J_H^y$. Both FM2-FM1 and FM1-FM0 transitions are accompanied by a change of the Fermi surface topology. For  relatively larger $\Delta_x/\Delta_y$, the FM2-FM1 and FM1-HFL transitions (with the light green area collapsing into a first-order transition line in reality) resemble the experimental observations in UGe$_2$ under pressure as reproduced in Fig. \ref{fig:Global}c, in particular the observed Fermi surface changes at the two transitions \cite{Terashima2001, Haga2002} and the magnetization plateau in between \cite{Pfleiderer2002}. For very large $\Delta_x/\Delta_y$ (inset), where the HFL covers the whole FM1 region and penetrates into the FM2 phase, one might find a first-order transition directly between FM2 and HFL. The FM phase transition in URhAl under pressure may correspond to this type  \cite{Combier2013,Shimizu2015}.

\begin{figure}
\centering\includegraphics[scale=0.42]{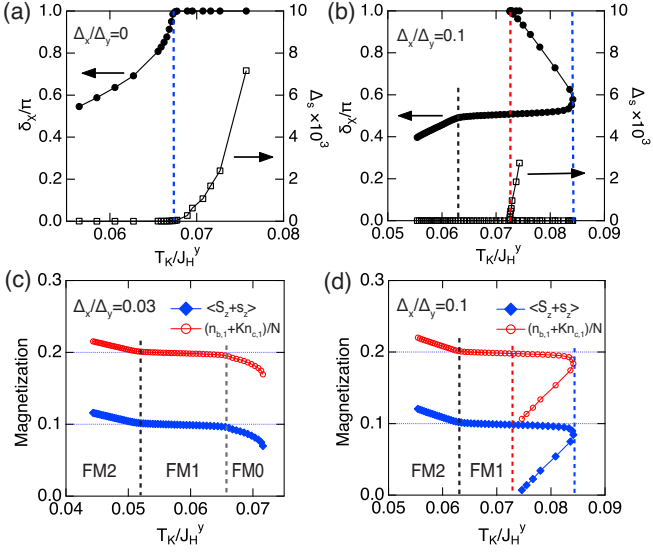}
\caption{\textbf{Holon phase shift and magnetization.} (a) The holon phase shift ($\delta_\chi$) and the spinon gap ($\Delta_s$) as functions of $T_K/J_H^y$ for $\Delta_x/\Delta_y=0$. (b) Same plot as (a), but for $\Delta_x/\Delta_y=0.1$.  Multiple solutions exist between the red and blue dashed lines. (c) Evolution of the total magnetization $\left\langle S_z+s_z \right\rangle$ and the quantity $(n_{b,1}+K n_{c,1})/N$ with respect to $T_K/J_H^y$ inside the FM phases. The parameters used are $\Delta_x/\Delta_y=0.03$, $\kappa=2S/N=0.2$ and $n_c/N=0.5$. (d) Same plot as (c), but for $\Delta_x/\Delta_y=0.1$.}
\label{fig:Mag}
\end{figure}

\begin{figure*}
\centering\includegraphics[scale=0.48]{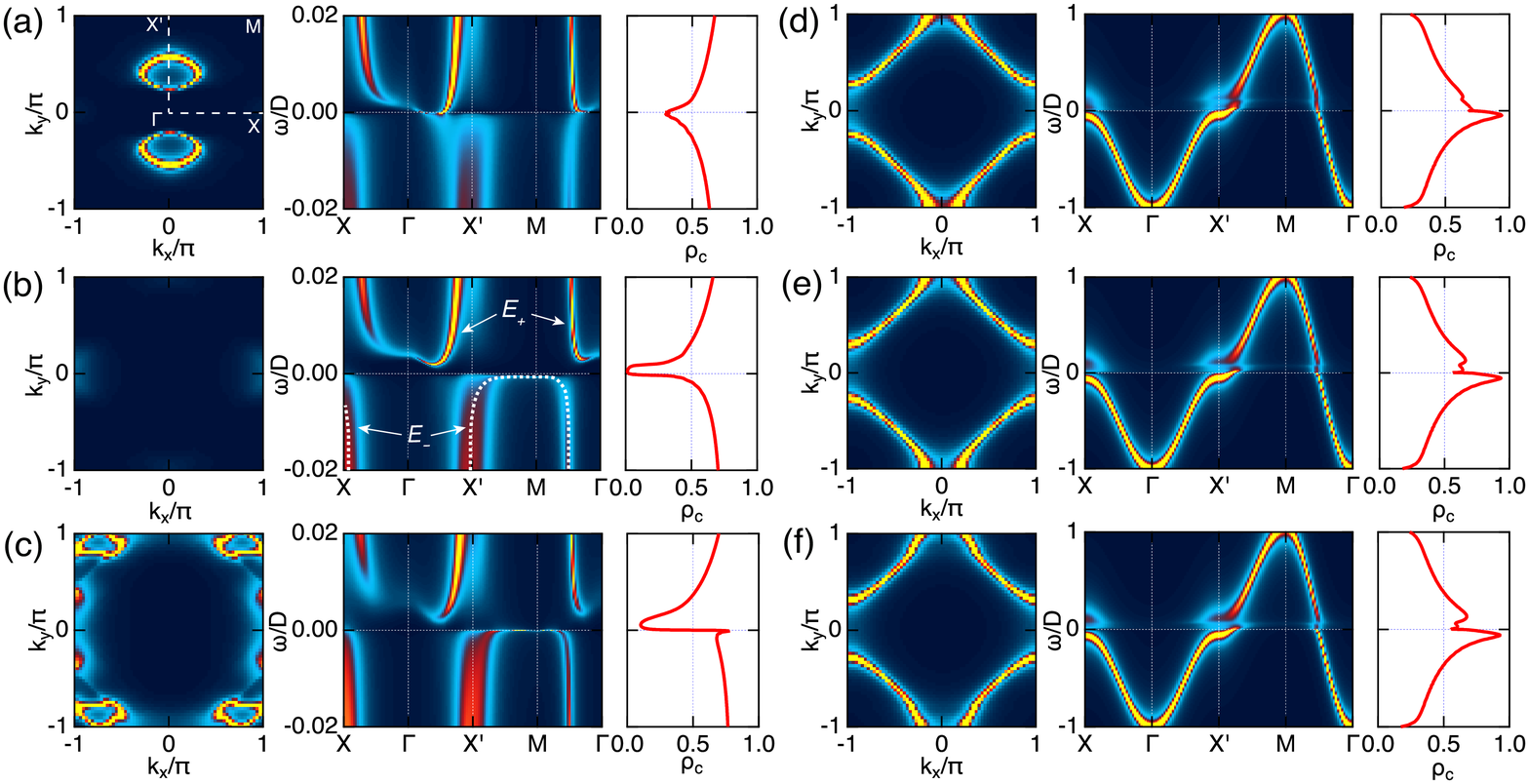}
\caption{\textbf{Fermi surface evolution with $T_K/J_H^y$.} (a)-(c) The conduction electron Fermi surface, dispersion, and density of states for the $\alpha=1$ component, at $\Delta_x/\Delta_y=0.03$ and different $T_K/J_H^y$. Two quasiparticle bands can be observed, denoted as $E_+$ and $E_-$. The dashed curve in (b) is a guide to the eye. (d)-(f) Same plots as (a)-(c), but for the $\alpha=2,\cdots, N$ components. A finite $N=4$ is taken to include the conduction electron self-energy in (d)-(f). The values of $T_K/J_H^y$ are: (a)(d) 0.044 (FM2 phase); (b)(e) 0.061 (FM1 phase); (c)(f) 0.071 (FM0 phase).}
\label{fig:FS}
\end{figure*}

To see more details on these different FM phases, we plot in Fig. \ref{fig:Mag} various quantities as functions of $T_K/J_H^y$ for different anisotropy ratios. The holon phase shift is defined as $\delta_\chi\equiv -\frac{1}{\mathcal{V}}\sum_{\bm{p}}\text{Im}\ln [-G_\chi(\bm{p},0)^{-1}]$. The quantity $\delta_\chi/\pi$ measures the fraction of local spins being effectively Kondo screened, which reaches unity in the HFL phase, indicating full Kondo screening. The spinon gap $\Delta_s$ is determined by the renormalized spinon energy at the $\Gamma$ point, satisfying $\Delta_s-\varepsilon_{{\Gamma}}^b-\text{Re}\Sigma_b(\Gamma,\Delta_s)=0$.  In the quasi-one dimensional case $\Delta_x/\Delta_y\rightarrow 0$, $\delta_\chi/\pi$ decreases continuously from unity at the FM0-HFL transition point, marked by the blue dashed line in Fig. \ref{fig:Mag}a. Right at this point, $\Delta_s$ vanishes and the spin susceptibility diverges (see Supplementary Materials), indicating formation of the FM long range order. We have thus a continuous FM-HFL transition and a FM QCP as observed in CeRh$_6$Ge$_4$ \cite{CeRhGe_nature}. The situation for $\Delta_x/\Delta_y=0.1$ is significantly different. Within a certain range of $T_K/J_H^y$ between the red and blue dashed lines of Fig. \ref{fig:Mag}b, both the HFL and FM solutions can exist at zero temperature, so the true ground state should correspond to a first-order transition with a jump of $\delta_\chi/\pi$ from $1$ to $0.5$ somewhere inside the coexisting region. The constant $\delta_\chi=0.5\pi$ is associated with the half-metal nature of the FM1 phase, as will be explained later. Further reducing $T_K/J_H^y$ drives the holon phase shift away from the $0.5\pi$ plateau and the system enters the FM2 phase for small $T_K/J_H^y$.

The distinctions between the three FM phases can be further elucidated by comparing their total magnetization $M_z=\left\langle S_z +s_z \right\rangle$, shown in Fig. \ref{fig:Mag}c and \ref{fig:Mag}d for $\Delta_x/\Delta_y=0.03$ and $0.1$,  respectively. We see that $M_z$ decreases with increasing $T_K/J_H^y$ in the FM2 phase, but keeps constant in the FM1 phase. In contrast to the wide coexisting region at $\Delta_x/\Delta_y=0.1$, there exists a stable FM0 phase next to the FM1 phase at $\Delta_x/\Delta_y=0.03$, where $M_z$ further decreases with increasing $T_K/J_H^y$.  We find that the constant magnetization of FM1 satisfies $M_z=\kappa(1-n_c/N)$, where $n_c=\sum_\alpha n_{c,\alpha}$ is the electron concentration per channel (see Fig. S3 of Supplementary Materials for different $\kappa$). As shown in Figs. \ref{fig:Mag}c and \ref{fig:Mag}d, our numerical calculations also find another relation of the FM1 phase, $n_{b,1}+Kn_{c,1}=2S$, which is analytically related to the magnetization plateau (see Supplementary Materials Section IV). The physical meaning of this identity can be understood more directly in the case of $N=2$ and $K=2S=1$, where it becomes $n_{b,\uparrow}+n_{c,\uparrow}=1$. This leads to the commensurability $n_{c,\uparrow}=n_{b,\downarrow}$ due to the constraint $n_{b,\uparrow}+n_{b,\downarrow}=2S=1$. Since $c_{\uparrow}$ and ${b_\downarrow}$ are both minority components in the FM phase, they can fully participate in the Kondo spin-flip scattering and the commensurability implies a full screening of this spin component. Meanwhile, the remaining majority components ($c_{\downarrow}$, ${b_\uparrow}$) are not involved in the Kondo scattering and contribute to the plateau in the total magnetization. 

These different screening properties have important consequences on their respective band structures, which can be best seen in Fig. \ref{fig:FS} showing the evolution of the electron Femi surface and dispersion relation inside the FM phases. Due to the hybridization between conduction electrons and holons, there exist two quasiparticle bands for the minority component ($\alpha=1$), which we denote as $E_{\pm}(\bm{p})$. Inside the FM2 phase (see Fig. \ref{fig:FS}a), the chemical potential intersects with the $E_+(\bm{p})$ band, so the Fermi surface contains two electron-like pockets. Inside the FM0 phase (see Fig. \ref{fig:FS}c), the chemical potential intersects with the $E_-(\bm{p})$ band, giving rise to four hole-like pockets in the Fermi surface. Right in between, the FM1 phase is featured with an indirect energy gap separating the $E_+(\bm{p})$ and $E_-(\bm{p})$ bands, and the chemical potential lies just within the gap. We have thus a spin selective Kondo insulator \cite{ZhangGuangMing2010,Peters2012,Golez2013,Peters1D2012}, in which the holon phase shift can be calculated from the quasiparticle dispersions via $\delta_\chi/\pi=\frac{1}{\mathcal{V}}\sum_{\bm{p}}[1+\theta(-\varepsilon_{\bm{p}}^c)-\theta(-E_+(\bm{p}))-\theta(-E_-(\bm{p}))]=\frac{1}{\mathcal{V}}\sum_{\bm{p}}\theta(-\varepsilon_{\bm{p}}^c)$, giving the constant $\delta_\chi/\pi=0.5$ in Fig. \ref{fig:Mag}b. 

Qualitatively, the above evolution of $E_{\pm}(\bm{p})$ can be explained by the dispersion relation given by the hybridization between conduction electrons and holons near the Fermi energy:
\begin{eqnarray}
E_{\pm}(\bm{p})\approx\frac{1}{2}\left(\varepsilon_{\bm{p}}^c-\varepsilon_{-\bm{p}}^\chi \pm \sqrt{(\varepsilon_{\bm{p}}^c+\varepsilon_{-\bm{p}}^\chi)^2+4Z_{-\bm{p}}^\chi b^2}\right),
\label{eq:Epm}
\end{eqnarray}
where $\varepsilon_{\bm{p}}^\chi=Z_{\bm{p}}^\chi/J_K^*(\bm{p})$ is the holon dispersion without spinon condensation, and $Z_{\bm{p}}^\chi=[-\partial_\omega \text{Re}\Sigma_\chi(\bm{p},\omega)|_{\omega=0}]^{-1}>0$ is its quasiparticle residue. As $T_K/J_H^y$ increases, $\varepsilon_{\bm{p}}^\chi$ gradually evolves from being fully above the Fermi energy, to being entirely below the Fermi energy in the HFL phase. This is a general feature of Kondo systems that accounts for the stability of forming Kondo bound states \cite{Wang2021,Wang2021SpinCurrent}.  As a result, $-\varepsilon_{-\bm{p}}^\chi$, and hence $E_{\pm}(\bm{p})$, evolves in the opposite way, causing the  evolution of the Fermi surfaces from electron to hole pockets. In addition, because $\varepsilon_{\bm{p}}^\chi$ becomes increasingly flat approaching the HFL phase boundary, the quasiparticles become more and more heavy, giving rise to the sharp quasiparticle peak at the Fermi energy in the density of states shown in Fig. \ref{fig:FS}c.

For comparison, we also show the Green's function of conduction electrons for other spin components $\alpha=2,\cdots, N$, whose self-energies are the order of $O(1/N)$. As an example, we take $N=4$ and show its evolution in Figs. \ref{fig:FS}d-\ref{fig:FS}f with the same values of $T_K/J_H^y$ taken in Figs. \ref{fig:FS}a-\ref{fig:FS}c. Unlike that of the $\alpha=1$ component, the Fermi surfaces for $\alpha=2,\cdots, N$ do not disappear in the FM1 phase, revealing the half-metallic nature of the FM1 phase. On the other hand, we see that their Fermi surface also expands slightly with $T_K/J_H^y$, which may be attributed to the band bending caused by the precursor effect of Kondo resonance in the self-energy. Both Figs. \ref{fig:FS}c and \ref{fig:FS}f suggest that the Kondo effect has partly taken place in the FM0 phase, which explains its highly reduced magnetic moment. The FM0-to-HFL phase transition is therefore not of a Kondo-breakdown type. Rather, our theory predicts the coexistence of localized moments and hybridization effect within FM0, which may be responsible for the discrepancy between the measured Fermi surfaces and those from density functional theory (DFT) calculations with fully localized $f$-electrons \cite{CeRhGe_SciBull2021}.

\begin{figure}
\centering\includegraphics[scale=0.5]{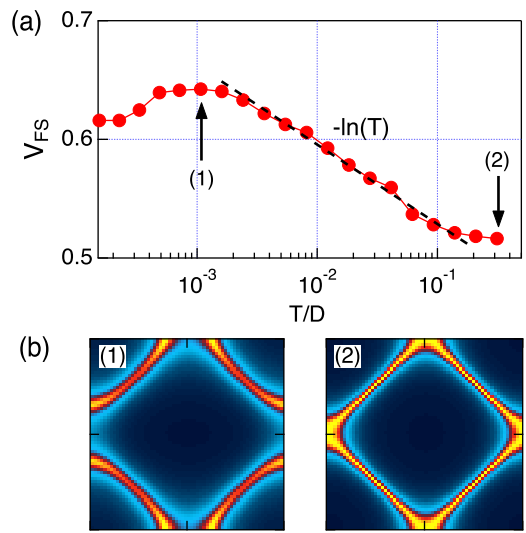}
\caption{\textbf{Fermi volume evolution with temperature.} (a) Temperature dependences of the electron Fermi volume ($V_{FS}=\frac{1}{\mathcal{V}}\sum_{\bm{p}}\theta(-\tilde{\varepsilon}_{\bm{p}}^c)$) above the FM0 ground state for $\Delta_x/\Delta_y=0.03$. The renormalized electron dispersion $\tilde{\varepsilon}_{\bm{p}}^c$ is determined by the electron Green's function with self-energy included at $N=4$. The dashed line indicates a $\ln(T)$ behavior of $V_{FS}$ at intermediate temperatures. (b) The electron Fermi surface at two different temperatures marked as (1) and (2) in (a).}
\label{fig:finiteT}
\end{figure}

To compare with spectroscopic measurements, we plot in Fig. \ref{fig:finiteT} the Fermi volume ($V_{FS}$) evolving with temperature above the FM0 ground state. At high temperatures, the Fermi volume is small and nearly half of the Brillouin zone, consistent with a half-filled conduction band with negligible Kondo effect. As the temperature decreases, the Fermi volume increases logarithmically, reaches a maximal value, and then decreases while approaching the FM0 ground state. The logarithmic increase of the Fermi volume indicates the development of Kondo hybridization at intermediate temperatures, consistent with the measurements of ARPES \cite{ARPES_PRL2021} and ultrafast optical spectroscopy \cite{pump_PRB2021}, while its decrease above the FM0 ground state results from a suppression of the Kondo hybridization due to enhanced spin fluctuations, confirming the  ``relocalization" phenomenon observed in many heavy fermion materials \cite{YangPNAS2017,Shier2012}.

\section{Discussion and conclusion}

To summarize, we have predicted a phase diagram of the anisotropic FM Kondo-Heisenberg model with both continuous and first-order FM quantum phase transitions. Our results provide a unified explanation of the very rich and qualitatively different experimental observations in two different types of FM Kondo lattice compounds CeRh$_6$Ge$_4$ and UGe$_2$ with extremely large or relatively small magnetic anisotropy, respectively. More specifically, for large anisotropy, we find a continuous quantum phase transition between the FM and HFL phases, consistent with that observed in the quasi-one-dimensional CeRh$_6$Ge$_4$. As the magnetic anisotropy is reduced, we find a coexisting region of FM and HFL, implying a first-order FM-HFL phase transition as observed in UGe$_2$. In addition, there exist three different types of FM ordered phases, among which the intermediate FM1 phase shows half-metallic property and is identified as a spin selective Kondo insulator. Our findings suggest the important role of magnetic anisotropy on interpreting the experimental observations in FM Kondo lattice materials.

It should be noted, however, that our model studies have ignored some peculiar microscopic details in realistic materials. For example, CeRh$_6$Ge$_4$ has a strong magnetic easy-plane anisotropy, which corresponds to an XXZ type Heisenberg interaction \cite{CeRhGe_nature}. This easy-plane anisotropy might favor a triplet resonating-valence-bond state \cite{Coleman2020tRVB}, cause a small discontinuity of the holon phase shift at the FM QCP, and further reduce the magnetic moment in the FM phase \cite{CeRhGe_nature}. In UGe$_2$, the strong spin-orbit coupling can result in FM fluctuations of Ising character, which may be important for its superconducting properties \cite{UGe2_review}. In a more realistic study, one may also want to consider the variation of conduction electron concentration with external parameters, which might tune the FM1-FM2 transition into a first-order one \cite{Wysokinski2015} or alter the sign of the RKKY interaction \cite{Lacroix2015}. These details may result in some additional interesting features but will not change our overall theoretical picture. It is straightforward to extend our theory to include these details in future studies.

\section{Conflict of interest}
The authors declare that they have no conflict of interest.

\section{Achnowledgments}
This work was supported by the National Key Research and Development Program of China (Grant No. 2017YFA0303103), the National Natural Science Foundation of China (Grants No. 12174429, No. 11774401, and No. 11974397), and the Strategic Priority Research Program of the Chinese Academy of Sciences (Grant No. XDB33010100).

\section{Author contributions}
Yang YF conceived and supervised the project. Wang JF performed the calculations. Wang JF and Yang YF wrote the manuscript.

\end{document}


\title{A unified theory of ferromagnetic quantum phase transitions in heavy fermion metals\\
\vspace{0.2cm}
- Supplementary Materials -}
\author{Jiangfan Wang}
\affiliation{Beijing National Laboratory for Condensed Matter Physics, Institute of Physics,
Chinese Academy of Science, Beijing 100190, China}
\author{Yi-feng Yang}
\affiliation{Beijing National Laboratory for Condensed Matter Physics,  Institute of Physics, 
Chinese Academy of Science, Beijing 100190, China}
\affiliation{School of Physical Sciences, University of Chinese Academy of Sciences, Beijing 100190, China}
\affiliation{Songshan Lake Materials Laboratory, Dongguan, Guangdong 523808, China}

\maketitle

\subsection{I. Self-consistent equations}
Within the SU($N$) Schwinger boson representation, the anisotropic ferromagnetic Kondo lattice model (Eq. (1) in the main text) has the following Lagrangian:
\begin{eqnarray}
\mathcal{L}&=&\sum_{\bm{p}a\alpha}c_{\bm{p}a\alpha}^\dagger (\partial_\tau+\varepsilon_{\bm{p}}^c)c_{\bm{p}a\alpha}+\sum_{\bm{p}\alpha}b_{\bm{p}\alpha}^\dagger (\partial_\tau+\varepsilon_{\bm{p}}^b)b_{\bm{p}\alpha}+\sum_{\bm{p}a}\frac{|\chi_{\bm{p}a}|^2}{J_K} \notag \\
& &+\frac{1}{\sqrt{\mathcal{V}N}}\sum_{\bm{pk}a\alpha}b_{\bm{p}\alpha}^\dagger c_{\bm{k}a\alpha}\chi_{\bm{p}-\bm{k},a}+c.c.+\mathcal{V}N\left( \frac{\Delta_x^2}{J_H^x}+  \frac{\Delta_y^2}{J_H^y}\right)-2\mathcal{V}\lambda S, 
\label{eq:L}
\end{eqnarray}
where $\varepsilon_{\bm{p}}^b=\lambda-2(\Delta_x \cos p_x+\Delta_y \cos p_y)$ is the bare spinon dispersion, and $\lambda$ is the Lagrange multiplier associated with the spinon constraint. The $\tau$-dependence of each field is implied. In the paramagnetic phase, the Green's functions of spinon, holon and conduction electron are:
\begin{eqnarray}
G_b(\bm{p},i\nu_n)&=&\frac{1}{i\nu_n-\varepsilon_{\bm{p}}^b-\Sigma_b(\bm{p},i\nu_n)}, \notag \\
G_\chi(\bm{p},i\omega_n)&=&\frac{1}{-J_K^{-1}-\Sigma_\chi(\bm{p},i\omega_n)}, \notag \\
G_c(\bm{p},i\omega_n)&=&\frac{1}{i\omega_n-\varepsilon_{\bm{p}}^c-\Sigma_c(\bm{p},i\omega_n)}.
\label{eq:G}
\end{eqnarray}
Both $\Sigma_b$ and $\Sigma_\chi$ are order of unity, while $\Sigma_c$ is order of $1/N$:
\begin{equation}
\Sigma_c(\bm{p},i\omega_n)=\frac{1}{N}\sum_{\bm{k}m}G_\chi(\bm{k}-\bm{p},i\nu_m-i\omega_n)G_b(\bm{k},i\nu_m).
\end{equation}
Therefore, one can ignore $\Sigma_c$ in the large-$N$ limit, and simply use the bare form of $G_c$ in the self-energy equations (Eq. (3) in the main text). At finite-$N$, one needs to include $\Sigma_c$ to calculate the renormalized dispersion and Fermi surface of conduction electrons. Minimizing the free energy with respect to $\lambda$ and $\Delta_\delta$ gives rise to the following constraints:
\begin{eqnarray}
\kappa&=&-\frac{1}{\beta\mathcal{V}}\sum_{\bm{p}n}G_b(\bm{p},i\nu_n), \notag \\
\frac{\Delta_\delta}{J_H^\delta}&=&-\frac{1}{\beta\mathcal{V}}\sum_{\bm{p}n}G_b(\bm{p},i\nu_n)\cos p_\delta . 
\label{eq:Con}
\end{eqnarray}
In practice, we choose different values of $\Delta_\delta$ as input parameters, and determine the value of $J_H^\delta$ using the second constraint of Eq. (\ref{eq:Con}) after self-consistency  has been achieved.

To study the FM ordered phase, we allow the spinons to have a static uniform condensation on the $\alpha=1$ spin component:
\begin{eqnarray}
b_{i\alpha}(\tau)\rightarrow \left\{ 
\begin{array}{rcl}
\sqrt{N}b & &\alpha=1 \\
b_{i\alpha}(\tau) & & \alpha=2,\cdots, N
\end{array}
\right..
\end{eqnarray}
The Lagrangian can then be separated into two parts, $\mathcal{L}=\mathcal{L}_1+\mathcal{L}_{N-1}$, with
\begin{eqnarray}
\mathcal{L}_1&=&\sum_{\bm{p}a}(c_{\bm{p}a,1}^\dagger, \chi_{-\bm{p},a})\left(\begin{array}{rcl} &\partial_\tau+\varepsilon_{\bm{p}}^c &  -b  \\ & -b &  -J_K^{-1} \end{array}\right)\left(\begin{array}{rcl}c_{\bm{p}a,1} \\ \chi_{-\bm{p},a}^\dagger \end{array} \right)+\mathcal{V}Nb^2(\lambda-2(\Delta_x+\Delta_y)),\notag \\
\mathcal{L}_{N-1}&=&\sum_{\bm{p}a,\alpha >1}c_{\bm{p}a\alpha}^\dagger (\partial_\tau+\varepsilon_{\bm{p}}^c)c_{\bm{p}a\alpha}+\sum_{\bm{p},\alpha>1}b_{\bm{p}\alpha}^\dagger (\partial_\tau+\varepsilon_{\bm{p}}^b)b_{\bm{p}\alpha}+\frac{1}{\sqrt{\mathcal{V}N}}\sum_{\bm{pk}a,\alpha>1}b_{\bm{p}\alpha}^\dagger c_{\bm{k}a\alpha}\chi_{\bm{p}-\bm{k},a}+c.c. \notag \\
& &+\mathcal{V}N\left( \frac{\Delta_x^2}{J_H^x}+  \frac{\Delta_y^2}{J_H^y}\right)-2\mathcal{V}\lambda S.
\end{eqnarray}
The condensation provides a hybridization field for the holons and conduction electrons of the $\alpha=1$ spin flavor, with  the following Green's functions:
\begin{eqnarray}
G_\chi(\bm{p},i\omega_n)&=&\frac{1}{-J_K^{-1}-\Sigma_\chi(\bm{p},i\omega_n)+\frac{b^2}{-i\omega_n-\varepsilon_{-\bm{p}}^c}}, \notag \\
G_c^{\alpha=1}(\bm{p},i\omega_n)&=&\frac{1}{i\omega_n-\varepsilon_{\bm{p}}^c+\frac{b^2}{-J_K^{-1}-\Sigma_\chi(-\bm{p},-i\omega_n)}}.
\label{eq:FG}
\end{eqnarray}
The Green's functions  of the Schwinger boson and conduction electron of the $\alpha>1$ spin flavors remain the same as in Eq. (\ref{eq:G}). Minimizing the free energy with respect to $\lambda$, $\Delta_\delta$ and $b$ gives the following constraints:
\begin{eqnarray}
\kappa&=&b^2-\frac{N-1}{N}\frac{1}{\beta\mathcal{V}}\sum_{\bm{p}n}G_b(\bm{p},i\nu_n),\notag \\
\frac{\Delta_\delta}{J_H^\delta}&=&b^2-\frac{N-1}{N}\frac{1}{\beta\mathcal{V}}\sum_{\bm{p}n}G_b(\bm{p},i\nu_n)\cos p_\delta, \notag\\
0&=&\lambda-2(\Delta_x+\Delta_y)+\text{Re}\Sigma_b(\bm{0},0),
\end{eqnarray}
where $(N-1)/N\rightarrow 1$ in the large-$N$ limit. The last constraint requires that the renormalized spinon spectrum touches zero energy at the $\Gamma$ point, consistent with our assumption of a uniform condensation. Again, only the first and the third constraints are solved self-consistently, while the second constraint is used to find $J_H^\delta$ at the end of calculation.

\subsection{II. The effects of biquadratic spin interaction term}
\begin{figure}
\centering\includegraphics[scale=0.45]{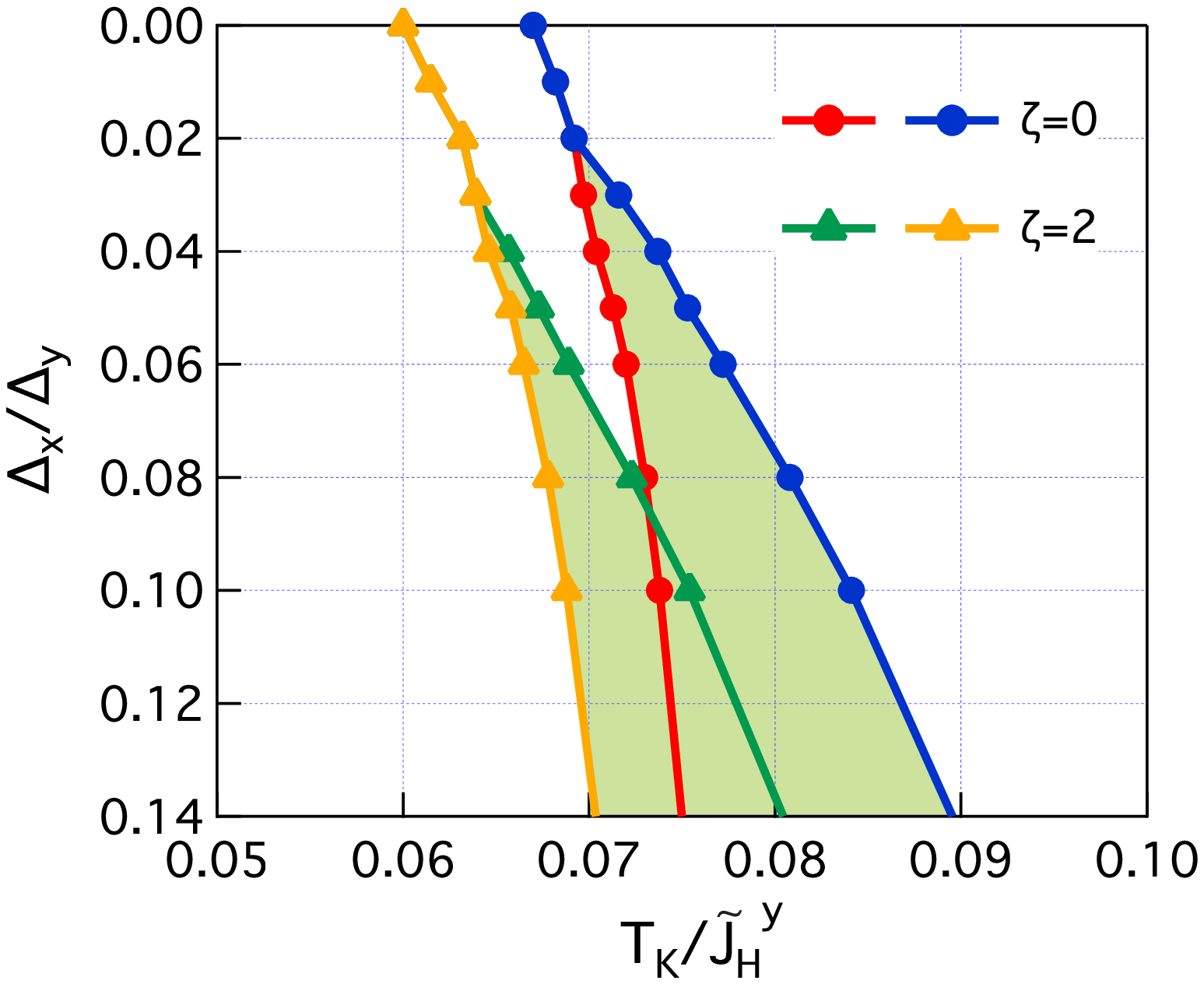}
\caption{\textbf{Effects of the biquadratic  term.} The phase diagram with ($\zeta=2$) or without ($\zeta=0$) the biquadratic term. The shadowed areas denote the coexisting regions. }
\label{fig:PDzeta}
\end{figure}

In order to alleviate the artificial first-order transition of the Schwinger boson mean-field theory, we introduce a biquadratic spin interaction term into the Hamiltonian, so that the ``Heisenberg'' term becomes 
\begin{eqnarray}
H_H&=&-\sum_{i\delta}J_H^\delta\left[\bm{S}_i\cdot \bm{S}_{i+\delta}+\zeta (\bm{S}_i\cdot \bm{S}_{i+\delta})^2\right] \notag \\
  &=& -\sum_{i\delta}J_H^\delta (1-\zeta/2)\bm{S}_i\cdot \bm{S}_{i+\delta}+C,
  \label{eq:BH}
\end{eqnarray}
where we have used the identity $(\bm{S}_i\cdot \bm{S}_{i+\delta})^2=3/16-(\bm{S}_i\cdot \bm{S}_{i+\delta})/2$, and $C$ is a constant. The condition that the effective Heisenberg coupling $J_H^\delta(1-\zeta/2)$ remains ferromagnetic requires $\zeta<2$. At large-$N$, the Heisenberg term is written as
\begin{equation}
H_H=-\sum_{i\delta}\frac{J_H^\delta}{N}\left(T_{i\delta}+\frac{2\zeta}{N^2}T_{i\delta}^2\right),
\end{equation}
where $T_{i\delta}=\sum_{\alpha\beta}S_i^{\alpha\beta}S_{i+\delta}^{\beta\alpha}$, and $S_{i}^{\alpha\beta}=b_{i\alpha}^\dagger b_{i\beta}-\kappa\delta_{\alpha\beta}$ is the SU($N$) generator. It reduces to Eq. (\ref{eq:BH}) in the case of $N=2$. Using the mean-field expansion $T_{i\delta}=\left\langle T_{i\delta}\right\rangle+\delta T_{i\delta}=\left\langle T_{i\delta}\right\rangle+(T_{i\delta}-\left\langle T_{i\delta}\right\rangle)$, one has
\begin{equation}
T_{i\delta}^2=-\left\langle T_{i\delta}\right\rangle^2+2\left\langle T_{i\delta}\right\rangle T_{i\delta}+O(\delta T_{i\delta}^2),
\end{equation}
so that
\begin{eqnarray}
H_H=-\sum_{i\delta}\frac{J_H^\delta}{N}\left(1+\frac{4\zeta}{N^2} \left\langle T_{i\delta}\right\rangle \right)T_{i\delta}+C,
\end{eqnarray}
where $C$ is a constant. The above equation defines a modified Heisenberg coupling
\begin{equation}
\tilde{J}_{H}^\delta\equiv J_H^\delta \left(1+\frac{4\zeta}{N^2} \left\langle T_{i\delta}\right\rangle \right)=J_H^\delta \left(1+\frac{4\zeta}{(\tilde{J}_{H}^\delta)^2} \Delta_\delta^2 \right),
\end{equation}
which suppresses the artificial first-order transition. Fig. \ref{fig:PDzeta} shows the phase diagram in terms of the modified Doniach ratio $T_K/\tilde{J}_H^y$ for $\zeta=0$ and $\zeta=2$.  The size of the coexisting region decreases as one includes the biquadratic term, and the tricritical point is shifted slightly from $\Delta_x/\Delta_y=0.02$ to $0.03$.

\subsection{III. Quantum criticality in the 1D limit}
\begin{figure}
\centering\includegraphics[scale=0.45]{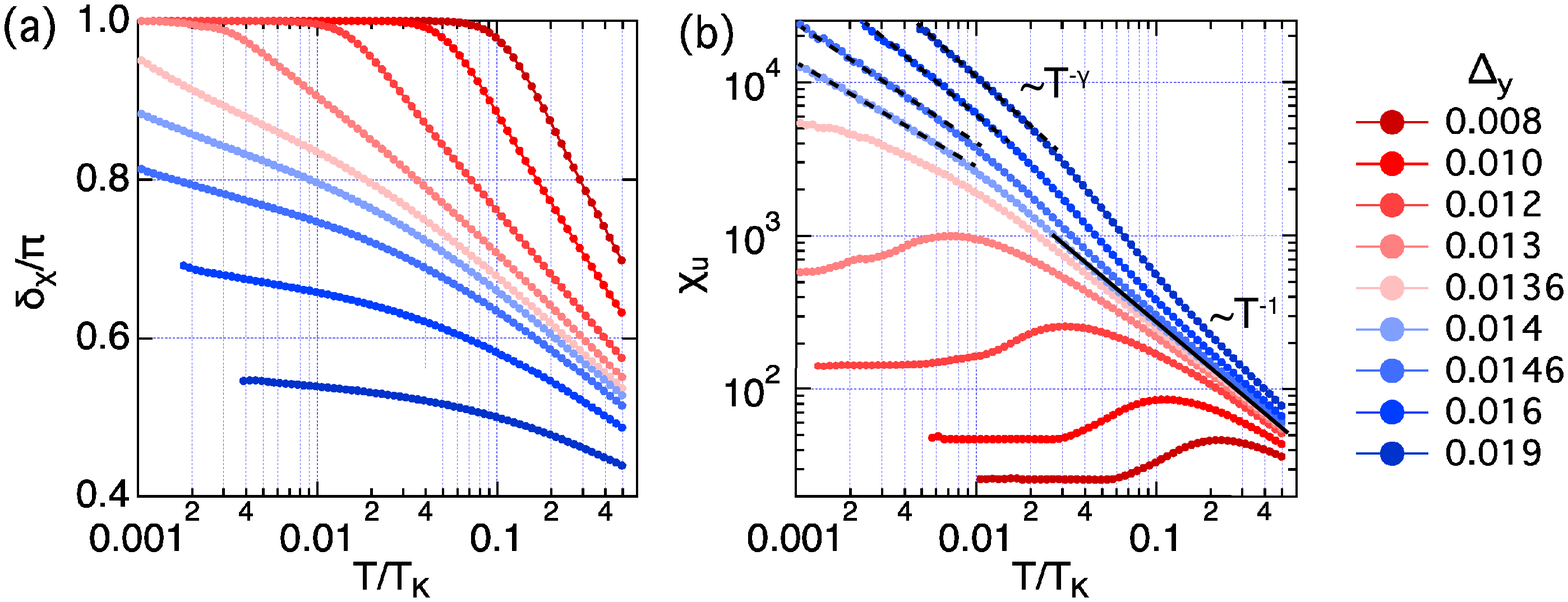}
\caption{\textbf{Quantum criticality at 1D limit.} Temperature dependence of (a) the holon phase shift and (b) the uniform impurity spin susceptibility at $\Delta_x=0$ and different $\Delta_y$. The black solid line indicates $T^{-1}$ Curie behavior at high temperature, while the dashed lines indicate $T^{-\gamma}$ divergence of the susceptibility at low temperature.  }
\label{fig:Phys}
\end{figure}

In Fig. \ref{fig:Phys}, we show physical quantities as functions of $T/T_K$ at $\Delta_x=0$ and different $\Delta_y$. The uniform susceptibility of impurity spin is calculated as
\begin{equation}
\chi_u=\frac{2}{\mathcal{V}}\sum_{\bm{k}}\int \frac{d\omega}{\pi}\frac{1}{e^{\beta \omega}-1}\text{Im}G_b(\bm{k},\omega)\text{Re}G_b(\bm{k},\omega).
\end{equation}
It follows the Curie law $T^{-1}$ at high temperatures, becomes constant in the heavy Fermi liquid phase (red) at low temperatures, and diverges with another power law $T^{-\gamma}$ on the ferromagnetic side of the QCP (blue). The QCP locates at $\Delta_y=0.014$ (or $T_K/J_H^y=0.067$), with a critical power index $\gamma \approx 0.67$. The power index $\gamma$ increases with $\Delta_y$ on the FM side due to the weakening of Kondo screening effect, as suggested by the decreasing holon phase shift.

\subsection{IV. Derivation of some identities in FM1}
In the SU($N$) representation, we define the impurity spin polarization along the $z$ direction as
\begin{equation}
M_z^S\equiv \frac{1}{\mathcal{V}} \sum_i\left\langle S_z(i) \right\rangle \equiv \frac{1}{\mathcal{V}N}\sum_i\left(\sum_{\alpha\leq N/2}\left\langle b_{i\alpha}^\dagger b_{i\alpha}\right\rangle- \sum_{\alpha> N/2}\left\langle b_{i\alpha}^\dagger b_{i\alpha}\right\rangle \right).
\end{equation}
This definition correctly gives $M_z^S=0$ in the paramagnetic phase, since the spinon concentration  $n_{b,\alpha}\equiv \frac{1}{\mathcal{V}}\sum_i \left\langle b_{i\alpha}^\dagger b_{i\alpha}\right\rangle$ is identical for all spin flavors. In the FM phase, the $\alpha=1$ spin flavor develops condensation, while all the other spin flavors remain equivalent. This gives us
\begin{eqnarray}
M_z^S=\frac{1}{N}(n_{b,1}-n_{b,\alpha\neq 1})=\frac{1}{N}(Nb^2-O(1))=b^2+O(1/N).
\end{eqnarray}
The spin polarization of conduction electrons can be defined in the same way
\begin{eqnarray}
m_z^s&\equiv& \frac{1}{\mathcal{V}N}\sum_{ia}\left(\sum_{\alpha\leq N/2} \left\langle c_{ia\alpha}^\dagger c_{ia\alpha}  \right\rangle -  \sum_{\alpha> N/2} \left\langle c_{ia\alpha}^\dagger c_{ia\alpha}  \right\rangle \right)\notag \\
&=& \frac{K}{N} (n_{c,1}-n_{c,\alpha\neq 1}),
\end{eqnarray}
where $n_{c,\alpha}\equiv\frac{1}{\mathcal{V}}\sum_i \left\langle c_{ia\alpha}^\dagger c_{ia\alpha}  \right\rangle$ is the electron concentration of spin flavor $\alpha$, which is independent of the channel index. As discussed in the main text, the FM1 phase also satisfies another identity, $n_{b,1}+K n_{c,1}=2S$, which results in $K n_{c,1}=(N-1)n_{b,\alpha\neq 1}$ when combined with the constraint $\sum_\alpha n_{b,\alpha}=2S$. This consistently leads to the plateau of the total magnetization:
\begin{eqnarray}
M_z&\equiv& M_z^S+m_z^s=\frac{n_{b,1}+K n_{c,1}-n_{b,\alpha\neq 1}-K n_{c,\alpha\neq 1}}{N} \notag \\
&=&\frac{2S-\frac{K}{N-1}(n_{c,1}+(N-1) n_{c,\alpha\neq 1})}{N} \notag \\
&=&\frac{2S-\frac{K}{N-1}n_c}{N},
\label{eq:Mz}
\end{eqnarray}
where $n_c=\sum_\alpha n_{c,\alpha}$ is the total electron concentration per channel. In the large-$N$ limit, Eq. (\ref{eq:Mz}) reduces to $M_z=\kappa (1-n_c/N)$. 

We now derive the following relation in the FM phases:
\begin{equation}
\frac{\delta_\chi}{\pi}=\frac{1}{\mathcal{V}}\sum_{\bm{p}}[1+\theta(-\varepsilon_{\bm{p}}^c)-\theta(-E_+(\bm{p}))-\theta(-E_-(\bm{p}))].
\end{equation}
Using Eq. (\ref{eq:FG}), we have
\begin{eqnarray}
\ln [-G_\chi(\bm{p},\omega^+)^{-1}]=\ln [(J_K^{-1}+\Sigma_\chi(\bm{p},\omega^+))(\omega^++\varepsilon_{-\bm{p}}^c)+b^2]-\ln[\omega^++\varepsilon_{-\bm{p}}^c],
\end{eqnarray}
where $\omega^+\equiv \omega+i0^+$.  Near the Fermi energy, one has
\begin{equation}
\Sigma_\chi'(\bm{p},\omega^+)=\Sigma_\chi'(\bm{p},0^+)+\partial_\omega \Sigma_\chi'(\bm{p},\omega^+)|_{0}\omega,
\end{equation}
where $\Sigma_\chi'$ ($\Sigma_\chi''$) denotes the real (imaginary) part of self-energy. Defining the quasiparticle residue $Z_{\bm{p}}^\chi\equiv [-\partial_\omega \Sigma_\chi'(\bm{p},\omega^+)|_{0}]^{-1}$ and the holon's dispersion relation $\varepsilon_{\bm{p}}^\chi\equiv Z_{\bm{p}}^\chi(J_K^{-1}+\Sigma_\chi'(\bm{p},0))$, we have
\begin{equation}
\ln [-G_\chi(\bm{p},\omega^+)^{-1}]=\ln [(\omega-\varepsilon_{\bm{p}}^\chi-i Z_{\bm{p}}^\chi \Sigma_\chi''(\bm{p},\omega^+))(\omega^++\varepsilon_{-\bm{p}}^c)-Z_{\bm{p}}^\chi b^2]-\ln[\omega^++\varepsilon_{-\bm{p}}^c]-\ln [-Z_{\bm{p}}^\chi].
\end{equation}
At low temperature, one has $\lim_{\omega\rightarrow 0}\Sigma_\chi''(\bm{p},\omega)=0^-$ and $Z_{\bm{p}}^\chi>0$, therefore $\lim_{\omega\rightarrow 0}(-i Z_{\bm{p}}^\chi \Sigma_\chi''(\bm{p},\omega^+))=i0^+$. The pole of the holon Green's function determines the quasiparticle spectrum:
\begin{eqnarray}
& &0=(\omega-\varepsilon_{\bm{p}}^\chi)(\omega+\varepsilon_{-\bm{p}}^c)-Z_{\bm{p}}^\chi b^2 \notag\\
&\rightarrow & \omega=\frac{\varepsilon_{\bm{p}}^\chi-\varepsilon_{-\bm{p}}^c\pm\sqrt{(\varepsilon_{\bm{p}}^\chi+\varepsilon_{-\bm{p}}^c)^2+4Z_{\bm{p}}^\chi b^2}}{2}= -E_\mp(-\bm{p}).
\end{eqnarray}
Note that $E_\pm(\bm{p})$ is defined as the quasiparticle pole of $G_c^{\alpha=1}(\bm{p},\omega)$, which is exactly the pole of $G_\chi(\bm{p},\omega)$ after the particle-hole transformation $\bm{p}\rightarrow {-\bm{p}}$, $\omega\rightarrow -\omega$.  
Therefore, we have
\begin{eqnarray}
\frac{\delta_\chi}{\pi}&=&-\frac{1}{\pi \mathcal{V}}\sum_{\bm{p}}\text{Im}\ln [-G_\chi(\bm{p},0^+)^{-1}] \notag \\
&=&-\frac{1}{\pi\mathcal{V}}\sum_{\bm{p}}\left(\text{Im}\ln [E_+(\bm{p})+i0^+]+\text{Im}\ln [E_-(\bm{p})+i0^+]-\text{Im}\ln [\varepsilon_{\bm{p}}^c+i0^+] \right)+1 \notag \\
&=&\frac{1}{\mathcal{V}}\sum_{\bm{p}}[1+\theta(-\varepsilon_{\bm{p}}^c)-\theta(-E_+(\bm{p}))-\theta(-E_-(\bm{p}))],
\end{eqnarray}
where we have assumed the inversion symmetry $E_{\pm}(\bm{p})=E_{\pm}(-\bm{p})$, $\varepsilon_{\bm{p}}^c=\varepsilon_{-\bm{p}}^c$.

\subsection{V. The effects of $\kappa$}

\begin{figure}
\centering\includegraphics[scale=0.34]{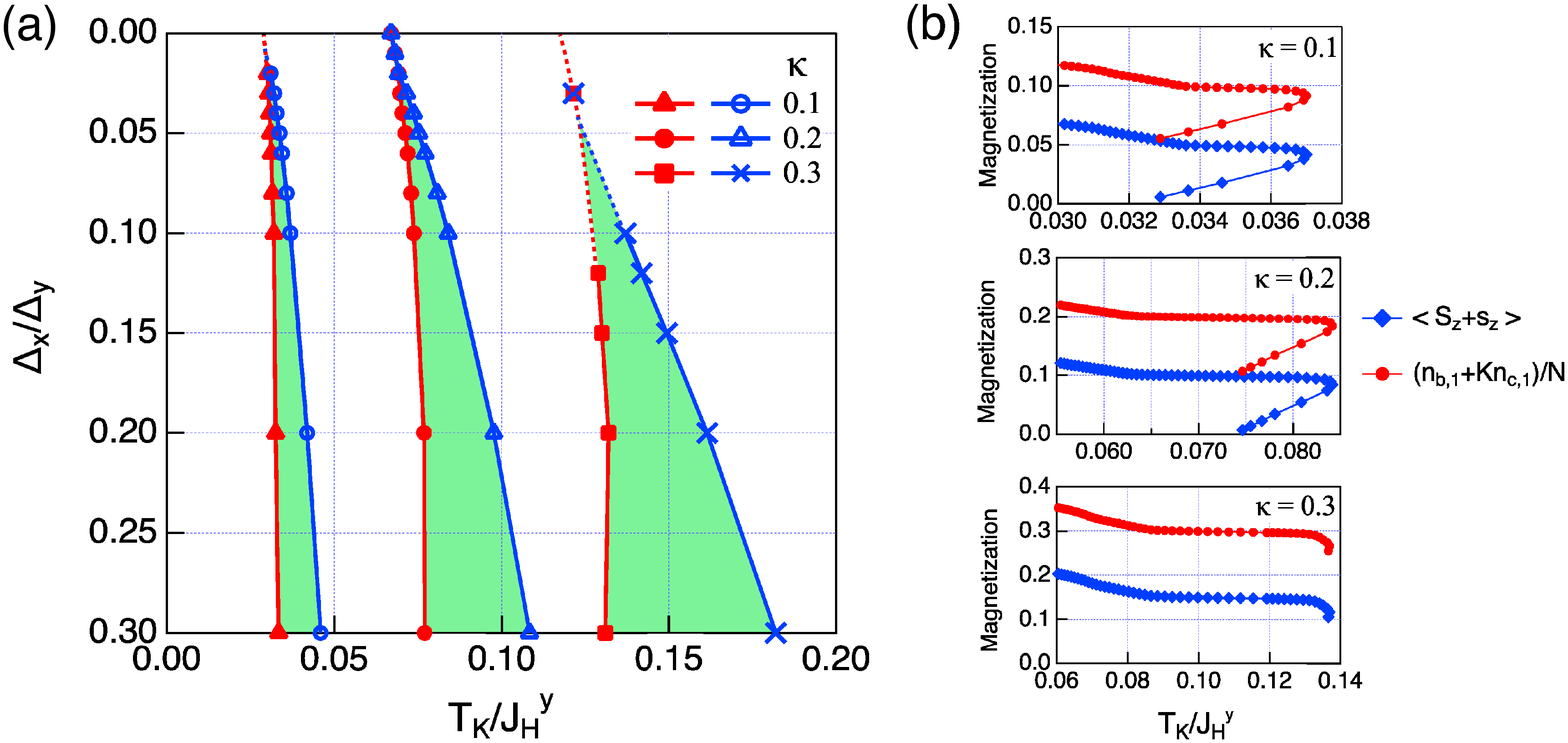}
\caption{\textbf{Effects of $\kappa$.} (a) Comparison of the coexisting region (light green areas) at different values of $\kappa$. The dashed lines are extrapolations. (b) The total magnetization and the quantity $n_{b,1}+Kn_{c,1}$ as functions of $T_K/J_H^y$ at $\Delta_x/\Delta_y=0.1$, $n_c/N=0.5$ and different values of $\kappa$.   }
\label{fig:kappa}
\end{figure}

The influence of $\kappa$ on the phase diagram is shown in Fig. \ref{fig:kappa}(a). For smaller $\kappa$, the FM phases are suppressed by the HFL phase, so that the coexisting region becomes narrower and is shifted towards smaller values of $T_K/J_H^y$. Roughly speaking, this is because a small value of $\kappa$ enhances the quantum zero motion of spin, thus suppresses the FM order and  favors the Kondo effect. We also find that the tricritical point shifts to larger values of $\Delta_x/\Delta_y$ as $\kappa$ becomes larger.  Fig. \ref{fig:kappa}(b) compares the plateaus of magnetization and the quantity $(n_{b,1}+Kn_{c,1})/N$ at different values of $\kappa$, with $\Delta_x/\Delta_y=0.1$ and $n_c/N=0.5$. One can see that the two relations $M_z=\kappa(1-n_c/N)$ and $(n_{b,1}+Kn_{c,1})/N=\kappa$ always hold inside the FM1 phase.
